\documentclass[a4paper,12pt]{article}
\usepackage{amssymb,amsfonts,amsmath,mathtext,cite,enumerate,float}
\usepackage{amssymb,amsthm,latexsym,euscript,amscd, authblk}
\usepackage{braket}

\usepackage[english]{babel}
\usepackage[cp1251]{inputenc}

\title {On errors generated by unitary dynamics of bipartite quantum systems}

\author[1]{G.G.Amosov\thanks {gramos@mi-ras.ru}}

\author[1]{A.S. Mokeev\thanks {alexandrmokeev@yandex.ru}}

\affil[1]{Steklov Mathematical Institute of Russian Academy of Sciences,
ul. Gubkina 8, Moscow 119991, Russia}

\begin{document} 

\maketitle

\begin {abstract}{ Given a quantum channel it is possible to define the non\-com\-mu\-tative operator graph whose properties determine a possibility of error-free transmission of information via this channel.
The corresponding graph has a straight definition through Kraus operators determining quantum errors. We are discussing the opposite problem of a proper definition of errors that some graph corresponds to. Taking into account that any graph is generated by some POVM we give a solution to such a problem by means of the Naimark dilatation theorem. Using our approach we construct errors corresponding to the graphs generated by unitary dynamics of bipartite quantum systems. 
The cases of POVMs on the circle group ${\mathbb Z}_n$ and the additive group $\mathbb R$ are discussed. As an example we
construct the graph corresponding to the errors generated by dynamics of two mode quantum oscillator.
  }
\end{abstract}

\section{Introduction} 

Since the famous work by Peter Shor \cite{Shor} it were developed many approaches to construction of quantum error-correcting codes. All these approaches require to have some conjectures on how noise acts on the quantum states which we want to preserve. Mathematically the action of a noise is described by a set of completely positive maps, called errors, on the set of states of a quantum system. The choice of this set is a separate theoretical problem. There are various types of conjectures, for example, Shor code \cite{Shor} corrects an arbitrary error in one physical qubit of a 9-qubit cluster in which we encode one logical qubit. Or in the case of stabilizer codes \cite{Gott} noise is given by a subset of $n$-fold Pauli group. 

In the general setting of that problem \cite{Knill1997} for any set of errors, it is possible to define a unique non-commutative operator graph \cite{Duan}, the knowledge of this graph is enough to determine the existence of quantum error-correcting code and define all possible codes for given errors. Such a correspondence between sets of errors and non-commutative operator graphs is not one-to-one, but each non-commutative operator graph describes codes for some set of errors in that sense\cite{Duan2, shirokov}. So it could be meaningful to explore the problem of quantum error-correcting codes for graphs with parametrizations given in some separate way from mentioned error-correction formalism, in that description each conjecture on a graph also is the conjecture on a possible noise. It is shown \cite{Duan2, shirokov, yashin}, that any non-commutative operator graph is the closure of the linear envelope of some positive operator valued measure (POVM). So the problem of quantum error correction could be viewed in terms of POVMs. In particular, if a non-commutative operator graph is linearly generated by a POVM covariant with respect to the action of a unitary-represented group it is possible to give the sufficient conditions of error-correcting code existence. Several examples of such graphs were constructed \cite{AM1,AM2, AM3, AMP}. 

Suppose that we have a solution to the error-correction problem for the graph which we define through some suitable parametrization, then it immediately appears a back problem on how to define errors generating this graph. In the present article we give partial solutions to this problem for graphs generated by POVMs. In section 2 we give a detailed description of the problem and describe the set of errors corresponding to the graph generated by POVM on the locally compact group $G$
by means of the Naimark dilation theorem. In section 3 we consider the special case of the cyclic group $G=\mathbb{Z}_n$. Section 4 is devoted to the group $G={\mathbb R}$. The example of two-mode quantum oscillator is studied in Section 5. The last section is for conclusions.

\section{Errors associated with graphs generated by POVM}

Denote $\mathfrak {S}_+(H)$ the set of positive nuclear operators in a Hilbert space $H$. Together with $\mathfrak {S}_+(H)$ we shall use the convex set of quantum states $\mathfrak {S}(H)$ consisting of unit trace operators from $\mathfrak {S}_+(H)$. Any $V$ belonging to the algebra of all bounded operators $B(H)$ in $H$
defines a linear completely positive map on $\mathfrak {S}_+(H)$ by the formula
\begin{equation}\label{er}
\Phi _V:\rho \to V\rho V^*,\ \rho \in \mathfrak {S}_+, 
\end{equation}
which can be considered as an error appearing  under the information transmission. In the case, a state $\rho \in \mathfrak {S}(H)$ is mapped to
$\frac {\Phi _V(\rho)}{Tr(V^*V\rho )}$.
Let us pick up the set of errors $\Phi _{V_j}$ having the form (\ref {er}) and define a linear space consisting of linear bounded operators
\begin{equation}\label{graph}
{\mathcal V}=\overline {span}(V_j^*V_k).
\end{equation}
Following to the general theory of error correcting codes \cite {Knill1997} 
if there exists the orthogonal projection $P$ with the property
\begin{equation}\label{P}
PV_j^*V_kP=c_{jk}P
\end{equation}
for some $c_{jk}\in \mathbb C$, then any error of the form (\ref {er}) for $V\in {\mathcal V}$ can be corrected if $supp\rho  \in PH$ in the sense that there is a completely positive map $\Psi $ such that
\begin{equation}\label{Phi}
\Psi \circ \Phi _V(\rho )=d_{\rho }\rho ,
\end{equation}
where $d_{\rho }>0$.

Given a quantum channel (completely positive trace preserving map) $\Phi :\mathfrak {S}(H)\to \mathfrak {S}(K)$ 
consider the Kraus decomposition
\begin{equation}\label{Kraus}
\Phi (\rho )=\sum \limits _{k}V_k\rho V_k^*,\ 
\end{equation}
where $V_k:H\to K$ are linear bounded operators and
\begin{equation}\label{Kraus2}
\sum \limits _kV_k^*V_k=I.
\end{equation}
A linear operator subspace $\mathcal {V}\subset B(H)$ defined by the formula
$$
{\mathcal V}=\overline {span}(V_j^*V_k)
$$
is said to be a non-commutative operator graph associated with $\Phi $. It immediately follows from the definition that
\begin{itemize}

\item $I\in \mathcal {V}$;

\item if $V\in \mathcal {V}$ then $V^*\in \mathcal {V}$.

\end{itemize}
Hence $\mathcal V$ is an operator system in the sense of \cite{ChoiEffros}. Moreover any operator system is a non-commutative operator graph associated with some quantum channel \cite{Duan2, shirokov, yashin}.
If there is a projection $P$ called a (quantum) anticlique \cite {weaver} that has the property
\begin{equation}\label{anti}
P{\mathcal {V}P}=\{\mathbb {C}P\}
\end{equation}
then (\ref {P}) is satisfied and all errors generated by $V\in {\mathcal V}$ can be corrected in the sense of (\ref {Phi}).

Let $G$ be a locally compact group with the Haar measure $\mu$.  Denote $\mathfrak {B}(G)$ the $\sigma$-algebra generated by compact subsets of $G$. The map $B\in \mathfrak {B}(G) \to M(B)$ from $\mathfrak {B}(G)$ to the cone of all positive operators in a Hilbert space $H$ is said to be a normalized positive operator valued measure (POVM) if \cite{Hol, Holevo2012}
$$
M(\emptyset)=0,\ M(G)=I
$$
and
$$
M(\cup _jB_j)=\sum \limits _jM(B_j), \ \textit{for} \  B_j \cap B_k=\emptyset ,\ j\neq k,
$$
where the sum in the last equation converges in weak operator topology.
Any non-commutative operator graph $\mathcal V$ is generated by some POVM $M(B)$ such that
\begin{equation}\label{graph}
\mathcal {V}=\overline {span}(M(B),\ B\in \mathfrak {B}(G)).
\end{equation}
Following to the Naimark dilation theorem $H$ can be isometrically embedded into a Hilbert space $K$ such that
\begin{equation}\label{Naimark}
M(B)=P_HE(B)|_H,\ B\in \mathfrak {B}(G),
\end{equation}
where $E(B)$ is an orthogonal projection valued measure and $P_H$ is a projection in $K$ onto $H$.
Following to the Naimark construction it is possible to pick up $K$ generated by all $E(B)H$.
Consider a set of completely positive maps $\Phi _{E(B)}:B(H)\to B(K)$ defined by
\begin{equation}\label{ERR0}
\Phi _{E(B)}(\rho )=E(B)\rho E(B),\ \rho \in \mathfrak {S}(H).
\end{equation}
Denote $V_B=E(B)|_H:H\to K$.

{\bf Proposition 1.} {\it Errors of the form (\ref {ERR0}) generate the non-commutative graph (\ref {graph}) in the sense
$$
\mathcal {V}=\overline {span}(V_B^*V_{B'},\ B,B'\in \mathfrak {B}(G))
$$
}

Proof.

Note that $V_B^*=P_HE(B):K\to H$. Hence 
$$
V_B^*V_{B'}=P_HE(B)E(B')=P_HE(B\cap B')=M(B\cap B'),
$$
$B,B'\in \mathfrak {B}(G)$.

$\Box $

\section{The cyclic group $G=\mathbb {Z}_n$.}

Following to the ideas of \cite {turk} let us consider two mutually unbiased bases $(e_j)_{j=1}^{n-1}$ and $(f_j)_{j=0}^{n-1}$ in a finite dimensional Hilbert space $K,\ dimK=n$. Define an orthogonal projection valued measure $E$ on $\mathbb {Z}_n$ by the formula
$$
E(\{j\})=\ket {e_j}\bra {e_j},\ j\in \mathbb {Z}_n
$$
and two unitary representations ${\mathbb Z}_n\ni j\to U_j$ and ${\mathbb Z}_n\ni j\to \hat U_j$ by the formulae
$$
U_j=\sum \limits _{k=0}^{d-1}e^{\frac {2\pi i}{n}jk}\ket {e_k}\bra {e_k},\ j\in \mathbb {Z}_n,
$$
$$
\hat U_j=\sum \limits _{k=0}^{d-1}e^{\frac {2\pi i}{n}jk}\ket {f_k}\bra {f_k},\ j\in \mathbb {Z}_n.
$$
Then,
$$
\hat U_jE(\{k\})\hat U_j^*=E(\{k+j\}),\ j,k\in {\mathbb Z}_n,
$$
and $E$ is covariant with respect to the action $j\to \hat U_j$.

Take a unit vector $f=\frac {1}{\sqrt {n-1}} \sum \limits _{k=0}^ne_k$ and define the subspace $H$ by the condition $h\in H$ iff $(h,f)=0$. It immediately follows from Proposition 1 that the statement below holds true.

{\bf Corrolary 1.} {\it 
The graph
\begin{equation}\label{z}
{\mathcal V}=span(P_HU_j|_H,\ j\in {\mathbb Z}_n)
\end{equation}
is generated by the errors
$$
\rho \to U_j\rho U_j^*,\ \rho \in \mathfrak {S}(H). 
$$
On the other hand, (\ref {z}) is generated by the POVM
$$
M(\{j\})=P_HE(\{j\})|_H,\ j\in {\mathbb Z}_n,
$$
in $H$.
}

Let $\mathbb{T}=[0,2\pi)$ be the circle group with the operation $+/2\pi$.
Following to \cite {AM2} put $H=\mathbb{C}^{n}\otimes\mathbb{C}^{n}$ for $n\ge 2$ and denote $\ket{j,k},\ j,k\in {\mathbb Z}_n$ the orhonormal basis in $H$. Let us consider the following reducible unitary representation of $\mathbb{T}$
$$
\hat U_{\varphi}\ket{j,k}=e^{i\varphi j}\ket{j,k},\ \varphi \in \mathbb {T}.
$$
We also need the orthonormal basis $\ket{\eta_j^k},\ j,k \in \mathbb {Z}_n$ consisting of the generalized Bell states \cite {Bell}
$$
\ket{\eta_j^k}=\frac {1}{\sqrt {n}}\sum\limits_{s\in {\mathbb Z}_n}e^{\frac{2\pi i}{n}sj}\ket{s,s-k}.
$$
So
$$
\hat U_{\varphi}\ket{\eta_j^k}=\sum\limits_{s\in \mathbb {Z}_n}e^{i\left(\varphi+\frac{2\pi j}{n}\right)s}\ket{s,s-k}.
$$
Consider the orthogonal projections
$$
Q_j=\sum \limits _{k\in \mathbb {Z}_n}\ket {j\ j-k}\bra {j\ j-k},\ j\in {\mathbb G}. 
$$
Then \cite{AM3}, the non-commutative operator graphs 
$$
{\mathcal V}_j=span (\hat U_{\varphi }Q_j\hat U_{\varphi }^*,\ \varphi \in \mathbb {T})
$$
coincide and the graph ${\mathcal V}\equiv {\mathcal V}_j$ has the following unitary generators
\begin{equation}\label{repre}
U_j=\sum \limits _{k,l\in \mathbb {Z}_n}\ket {\eta _{k+j}^l}\bra{\eta _k^l},\ j\in \mathbb {Z}_n.
\end{equation}

Now Proposition 1 gives rise to the following.

{\bf Corrolary 2.} {\it Formula (\ref {repre}) determines a unitary representation of $\mathbb {Z}_n$ in $H$. The graph $\mathcal {V}=span(U_j,\ j\in {\mathbb Z}_n)$ is generated by the errors
$$
\rho \to U_j\rho U_j^*,\ \rho \in \mathfrak {S}(H),\ j\in \mathbb {Z}_n.
$$
}

\section{The case $G=\mathbb R$.}

Suppose that $H$ is isometrically embedded into $K=H\otimes H_E$ by means of the rule
\begin{equation}\label{emb}
f\to f\otimes e,\ f\in H,
\end{equation}
where $e$ is a fixed unit vector in a Hilbert space $H_E$.

Let $U_t:H\otimes H_E\to H\otimes H_E,\ t\in \mathbb R,$ be a one-parameter unitary group describing the interaction between the system $H$ and its environment $H_E$.
Denote $\mathfrak {B}({\mathbb R})$  the $\sigma$-algebra of Borel sets on the real line $\mathbb R$.
The Stone theorem reads
$$
U_t=\int \limits _{\mathbb R}e^{itx}E(dx),\ t\in {\mathbb R},
$$
where $E$ is an orthogonal projection valued measure on $\mathbb R$. 
Our goal is to protect information encoded by states belonging to $\mathfrak {S}(H)$ against errors having the form (\ref {ERR0}). Denote $\mathfrak {A}$ the commutative algebra generated by the projections $E(B),\ B\in \mathfrak {B}({\mathbb R})$. We consider linear completely positive maps from 
$\mathfrak {S}_+(H)$ to $\mathfrak {S}_+(H\otimes H_E)$ defined by the formula
\begin{equation}\label{er3}
\Phi _A(\rho )=A(\rho \otimes \rho _e)A^*,\ \rho \in \mathfrak {S}(H),\ A\in \mathfrak {A},\ \rho _e=\ket {e}\bra {e}\in \mathfrak {S}(H_E)
\end{equation}
as errors that can occur under the information transmission. 
Given $A\in \mathfrak {A}$ denote $V_A:H\to H\otimes H_E$ the linear operator defined by the formula
$$
V_Af=A(f\otimes e),\ f\in H,
$$
then the adjoint operator $V_A^*:H\otimes H_E\to H$.
Consider the linear operator space
\begin{equation}\label {ERR}
{\mathcal V}=\overline {span}(V_A^*V_{A'},\ A,A'\in \mathfrak {A})
\end{equation}

{\bf Proposition 2.} {\it Suppose that $\rho _e=\ket {e}\bra {e}\in \mathfrak {S}(H_E)$ is a fixed pure state of the environment. Then, the maps $\rho \to U_t(\rho \otimes \rho _e)U_t^*$ have the form (\ref {er3}). Moreover, the set of errors $\rho \to U_t\rho U_t^*,\ t\in {\mathbb R}$ generates the same operator space  as (\ref {ERR}).

}

{\bf Proof.}

Given a partition of $\mathbb R$ consisting of $B_j\in \mathfrak {B}({\mathbb R})$ such that $B_j\cap B_k=\emptyset,\ \cup _jB_j=\mathbb {R}$ and the corresponding projection resolution of identity $E_j=E(B_j)$
let us define the group of unitary operators 
$$
U_t^{(B_j)}=\sum \limits _{j=1}^ne^{its_j}E_j,\ t\in {\mathbb R},
$$
where $s_j\in B_j$. Then, $U_t^{(B_j)}\in \mathfrak {A}$. Hence, $U_t\in \mathfrak {A}$ also because $\mathfrak {A}$ is closed. To finish the proof it is sufficiently to take into account that $\overline {span}(U_t,\ t\in {\mathbb R})=\mathfrak {A}$.

$\Box $

\section{Example: two mode oscillator}

Here we study the explicit example described in \cite{AMP} in view of Section 4. For a Hamiltonian of a two-mode quantum oscillator
\begin{equation}\label{Hamilt}
\mathbf{H}=-\frac {1}{2}\frac {\partial ^2}{\partial x^2}-\frac {1}{2}\frac {\partial ^2}{\partial y^2}+\frac {(x-y)^2}{2},\ 
\end{equation}
acting in the bipartite quantum system $K=L^2({\mathbb R})\otimes L^2({\mathbb R})$,
it is possible to give formulae for the unitary group $U_t=e^{-it\mathbf{H}}$ in terms of the products of coherent states. Given a complex number $\alpha \in {\mathbb C}$, the coherent state is an eigenvector of the annihilation operator corresponding to eigenvalue $\alpha $. The wave function of a coherent state equals
$$
\xi_{\alpha}(x)=\frac{1}{\pi^{1/4}}exp\left(-\frac{|\alpha|^2}{2}\right)exp\left(-\frac{x^2-2\sqrt{2}\alpha x+\alpha^2}{2}\right),\  \alpha \in \mathbb{C}.
$$
Given two complex numbers $\alpha $ and $\beta $ let us consider the following products
$$
\psi _{\alpha\beta }(x,y)=\frac {1}{\sqrt {2}}\xi _{\alpha }\left(\frac{x+y}{\sqrt[4]{2}}\right)\xi _{\beta }\left(\frac{x-y}{\sqrt[4]{2}}\right).
$$
The group $U_t$ can be defined by its action on the overcomplete system $\psi _{\alpha\beta }$ such that
\begin{equation}\label{action}
(U_t\psi _{\alpha \beta})(x,y)=\frac {e^{-\frac {it}{\sqrt{2}}}}{\sqrt 2\sqrt {1+\sqrt{2}ti}}\xi _{\alpha }\left (\frac {x+y}{\sqrt[4]{2}\sqrt {1+\sqrt{2}ti}}\right ) \xi _{e^{-i\sqrt{2}t}\beta }\left(\frac{x-y}{\sqrt[4]{2}}\right).
\end{equation}
Take new variables
$$
\tilde x=\frac{x+y}{\sqrt[4]{2}},\ \tilde y=\frac{x-y}{\sqrt[4]{2}}
$$
and denote
$$
\braket {z|\alpha }=\frac {1}{\sqrt[4]{2}}\xi _{\alpha }(z), z\in {\mathbb C}. 
$$
Then, (\ref {action}) goes to 
\begin{equation}\label{action2}
\braket {\tilde x,\tilde y |U_t|\alpha \beta}=\frac {e^{-\frac {it}{\sqrt{2}}}}{\sqrt {1+\sqrt{2}ti}}\Braket {\frac {\tilde x}{\sqrt {1+\sqrt{2}ti}}|\alpha }\braket {\tilde y|e^{-i\sqrt{2}t}\beta }.
\end{equation}
Following to (\ref {emb}) let us introduce the isometrical embedding of $H=L^2({\mathbb R})$ into $K=L^2({\mathbb R})\otimes L^2({\mathbb R})$ by the formula
$$
\ket {\alpha }\to \ket {\alpha }\otimes \ket {\beta }\equiv \ket {\alpha \beta },
$$ 
where $\beta $ is a fixed complex number, while $\alpha $ runs over $\mathbb C$.

Put $\rho _e=\ket {\beta }\bra {\beta}$ and denote $\tilde H=\overline {span}(f\otimes \ket {\beta},\ f\in H)$. The following Corollary is due to Proposition 2.

{\bf Corollary 3.} {\it The set of errors
$$
\rho \to U_t(\rho \otimes \rho _e)U_t^*,\ t\in {\mathbb R},
$$
generates the non-commutative operator graph
$$
{\mathcal V}=\overline {span}(T_t=P_{\tilde H}U_t|_{\tilde H},\ t\in {\mathbb R}),
$$
where
$$
\braket {\tilde x,\tilde y|T_t|\alpha \beta }=exp\left(e^{-i\sqrt{2}t}|\beta|^{2}-\frac {it}{\sqrt{2}}\right) \left(1+\sqrt{2}ti\right)^{-\frac{1}{2}}\Braket {\frac {\tilde x}{\sqrt {1+\sqrt{2}ti}}|\alpha }\braket {\tilde y|\beta }.
$$
}

\section{Conclusion}

We discussed the problem of searching for a quantum noise corresponding to the operator graph. It is shown that if the graph is linearly generated by POVM, then the Naimark dilatation operators determine errors for our graph. The cases of the graphs generated by covariant POVMs on the cyclic group $\mathbb{Z}_n$ and on the additive group of reals $\mathbb{R}$ were discussed separately. Also we introduced the explicit example originated from the unitary dynamics of the two-mode quantum oscillator.

\section*{Acknowledgments} The work is supported by Russian Science Foundation under the grant no. 19-11-00086 and performed in Steklov Mathematical Institute of Russian Academy of Sciences.

\end{document}